\documentstyle[emulateapj,epsf]{article}
 
\begin{document}

\title{Upper Limits On Periodic, Pulsed Radio Emission from the X-Ray Point Source in Cassiopeia A}
\author{M.~A.~McLaughlin\altaffilmark{1}, J.~M.~Cordes\altaffilmark{1}, A.~A.~Deshpande\altaffilmark{2,3},
  B.~M.~Gaensler\altaffilmark{4,5},
T.~H.~Hankins\altaffilmark{2}, V.~M.~Kaspi\altaffilmark{6,7}, \& 
J.~S.~Kern\altaffilmark{2}}
\altaffiltext{1}{Astronomy Department, Cornell University, Ithaca, NY 14853}
\altaffiltext{2}{Physics Department, New Mexico Institute of Mining and 
Technology, Socorro, NM 87801}
\altaffiltext{3}{Raman Research Institute, Bangalore 560080, India}
\altaffiltext{4}{Hubble Fellow}
\altaffiltext{5}{Center for Space Research, Massachusetts Institute of Technology, Cambridge, MA 02139}
\altaffiltext{6}{Physics Department, McGill University, Montreal, Canada H3A2T8}
\altaffiltext{7}{on leave from Physics Department and Center for Space Research, Massachusetts Institute of Technology, Cambridge, MA 02139}
\begin{abstract}

The {\it Chandra X-ray Observatory} recently discovered an X-ray point source near the center
of Cassiopeia A, the youngest known Galactic supernova remnant.
We have conducted a sensitive search for radio pulsations from this source with the Very Large Array,
taking advantage of the high angular resolution of the array to resolve out the emission from the remnant
itself. 
No convincing signatures of a dispersed, periodic source or of isolated dispersed pulses
were found, whether for an isolated or a binary source.
We derive upper limits of 30 and 1.3 mJy  at 327 and 1435 MHz for its phase-averaged
pulsed flux density. The corresponding luminosity limits are lower than those
for any known radio pulsar with age less than $10^{4}$ years.
Our search sensitivities to single pulses were
25 and 1.0 Jy at 327 and 1435 MHz. For comparison, the Crab pulsar emits
roughly 80 pulses per minute with flux densities greater than 100 Jy at 327 MHz
and 8 pulses per minute with flux densities greater than
50 Jy at 1435 MHz. These limits suggest that Cas~A belongs to the growing population
of young neutron stars that are radio quiet 

\end{abstract}
\section{Introduction}

Cassiopeia~A, with an age of only 320 years, is the youngest known Galactic supernova
remnant. 
The supernova that gave rise to this remnant was probably discovered in 1680,
when Flamsteed (1725) observed a 6th magnitude star within $13'$ of Cas~A's location (\cite{ash80}).
Because its abundances of heavy elements are consistent with those expected from the explosion
of a massive star (\cite{hughes00}), Cas~A is thought to be the remnant of a Type II supernova.
Given the stellar initial mass function, a large fraction 
of Type II supernovae are expected to leave behind neutron
stars. While many searches at various wavelengths for the compact remnant in Cas~A have been undertaken,
only upper limits had been established until the recent {\it Chandra} discovery of an X-ray point source
(hereafter XPS)
near the center of the remnant (\cite{tan99}). This XPS was
later confirmed in archival {\it ROSAT} (\cite{asc99}) and {\it Einstein} (\cite{pav99}) images. The XPS is 
within $5''$ of the expansion center of the remnant (\cite{van83}). For a remnant distance of
$3.4^{+0.3}_{-0.1}$ kpc, calculated through radial velocity, proper motion, and age data (\cite{reed95}),
this offset corresponds to a maximum transverse velocity of roughly 200 km s$^{-1}$,
a typical velocity for young pulsars (\cite{cor98}).

Pavlov et al. (2000) find that the X-ray spectrum of the XPS is fit well by a power law with index
$2.6 - 4.1$, with luminosity $(2-60) \times 10^{34}$ ergs s$^{-1}$. Similarly,
Chakrabarty et al. (2001) estimate a power law index of $2.8 - 3.6$ and luminosity of
$(2 - 16) \times 10^{34}$ ergs s$^{-1}$. These indices are much
steeper than the X-ray spectra of Active Galactic Nuclei (AGN).
Furthermore, given the measured surface density of AGN, the probability of finding an
AGN that close to the center of the SNR
is negligible. For these reasons,
Pavlov et al. (2000) conclude that the XPS is almost certainly
associated with the Cas~A remnant. The fitted power law indices for the XPS are also
much steeper than those
measured
for the 6 young X-ray pulsars having ages less than $10^{4}$ years,
whose spectral indices range from $1.1-1.7$
(\cite{deepto01}). In addition, the estimated XPS 
luminosities are only marginally consistent with measured young pulsar X-ray luminosities,
which range from $3\times 10^{35}$ to $4\times 10^{36}$ ergs s$^{-1}$.
It is also significant that, unlike 5 of the 6
young pulsars with ages less than
$10^{4}$ years, Cas~A shows no evidence for a synchrotron nebula surrounding the point source. 

While there is much evidence for and against the neutron star nature of the XPS,
the definitive proof would be the 
detection of pulsations. No X-ray pulsations have been found in data taken with the {\it Chandra}
High Resolution Camera. However, a time-tagging problem caused the timing accuracy to be degraded from
16 $\mu$s to $\approx$ 4 ms, allowing searches for only broad pulses with periods greater than 20 ms 
(\cite{deepto01}). The detection of radio pulsations, however,
would be just as conclusive. For this reason, we have conducted a search for radio pulsations
from the XPS using the Very Large Array (VLA) at frequencies of 
327 and 1435 MHz. There have been several previous searches for a radio pulsar in Cas~A
(\cite{davies70,seir80,woan93,lorimer98}).
Our VLA search is many times more sensitive, as having an accurate position for the
XPS allows us to resolve out the strong emission from the remnant, which with 327- and 1435-MHz
 flux densities
of 6400 and 2000 Jy is one of the strongest radio sources
known.

\bigskip

\section{Data}

We observed the XPS on three days (23 October 1999, 31 October 1999,
and 28 November 1999) with the phased-array VLA in B array,
where the maximum baseline
is 11.4 km. The J2000 
coordinates of the XPS are RA=$23^{\rm h}23^{\rm m}27^{\rm s}.94$, DEC=$+58^{o}48'42''$,
with positional uncertainty of $1''$ (\cite{tan99}).
As the 327- and 1435-MHz synthesized beamwidths
are approximately 20 and 4 arcseconds respectively,  one beam at each frequency was sufficient
for this search. During each session, the XPS was observed at center frequencies of 327 and 1435 MHz,
resulting in three pointings of length 20, 40, and 50 minutes
at 327 MHz, and four pointings of length 40, 40, 45, and 50
minutes at 1435 MHz.
During each observing session,
we observed the known pulsar PSR B0355+54, which has a period of 154 ms and a dispersion measure
(DM) of 57 pc cm$^{-3}$.
A blank-sky pointing and the flux and phase calibrator J2350+646 (with
flux densities of 27 and 5 Jy at 327 and 1435 MHz) were also observed.

We used the VLA's High Time Resolution Processor (Moffett 1997) to record dual circular polarization data 
at a sampling frequency of 9.22 kHz. 
 Each scan was started on a 10-second tick, tied to the VLA's
hydrogen maser, compared to Universal Time through the GPS.
At 327 MHz, $14\times0.125$-MHz frequency channels covered a total bandwidth of 1.75 MHz.
These channels were spaced non-contiguously
to avoid known interference frequencies and covered a total bandwidth range of 3 MHz.
At 1435 MHz, $14\times4$-MHz channels covered a total contiguous bandwidth of 48 MHz. (As only 50 MHz of
bandwidth is available, some channels were duplicated). The gain in each of the 28 channels (2 
polarizations and 14 frequencies) was calculated from observations of the calibrator source.
 
\section{Analysis}

For each of the seven XPS data sets, we first summed right and left circular polarizations and then
dedispersed the data over a range of trial DMs. The Taylor and Cordes (1993) model predicts a
maximum DM of approximately 75 pc cm$^{-3}$ in the direction of the XPS. We dedispersed the data with
60 trial DMs
from 0 to 300 pc cm$^{-3}$ to allow for an underestimate or a contribution from Cas~A.
These 60 DMs were spaced nonuniformly, with
greater spacing at higher DMs, where the temporal smearing due to dispersion
across an individual frequency channel
is larger. Each dedispersed time series was Fourier transformed with 32 Mpoint
and/or 16 Mpoint Fast Fourier Transforms (FFTs), depending on the length of the data set.
Shorter FFTs of 8 and 4 Mpoints were done to increase our sensitivity
to some binary
pulsars. The resulting
power spectra were harmonically summed (up to a maximum of 16 harmonics) to increase our
sensitivity to narrow pulses (\cite{lyne98}). All features in the resulting power
spectra with signal-to-noise above a threshold of seven were saved as pulsar candidates.
For each candidate, the corresponding dedispersed time series was folded at the period corresponding to the
fundamental harmonic.
These profiles
were inspected and candidate lists from different days and at different frequencies
were compared. While many periodic signals were detected, they were predominantly at low DM,
had noiselike
pulse profiles, and did not appear on multiple epochs or at multiple frequencies, and were therefore
attributed to radio frequency interference (RFI).
Figure 1 shows the folded pulse profiles resulting from applying the search
algorithm to PSR B0355+54. 

\medskip
\epsfxsize=8truecm
\epsfbox{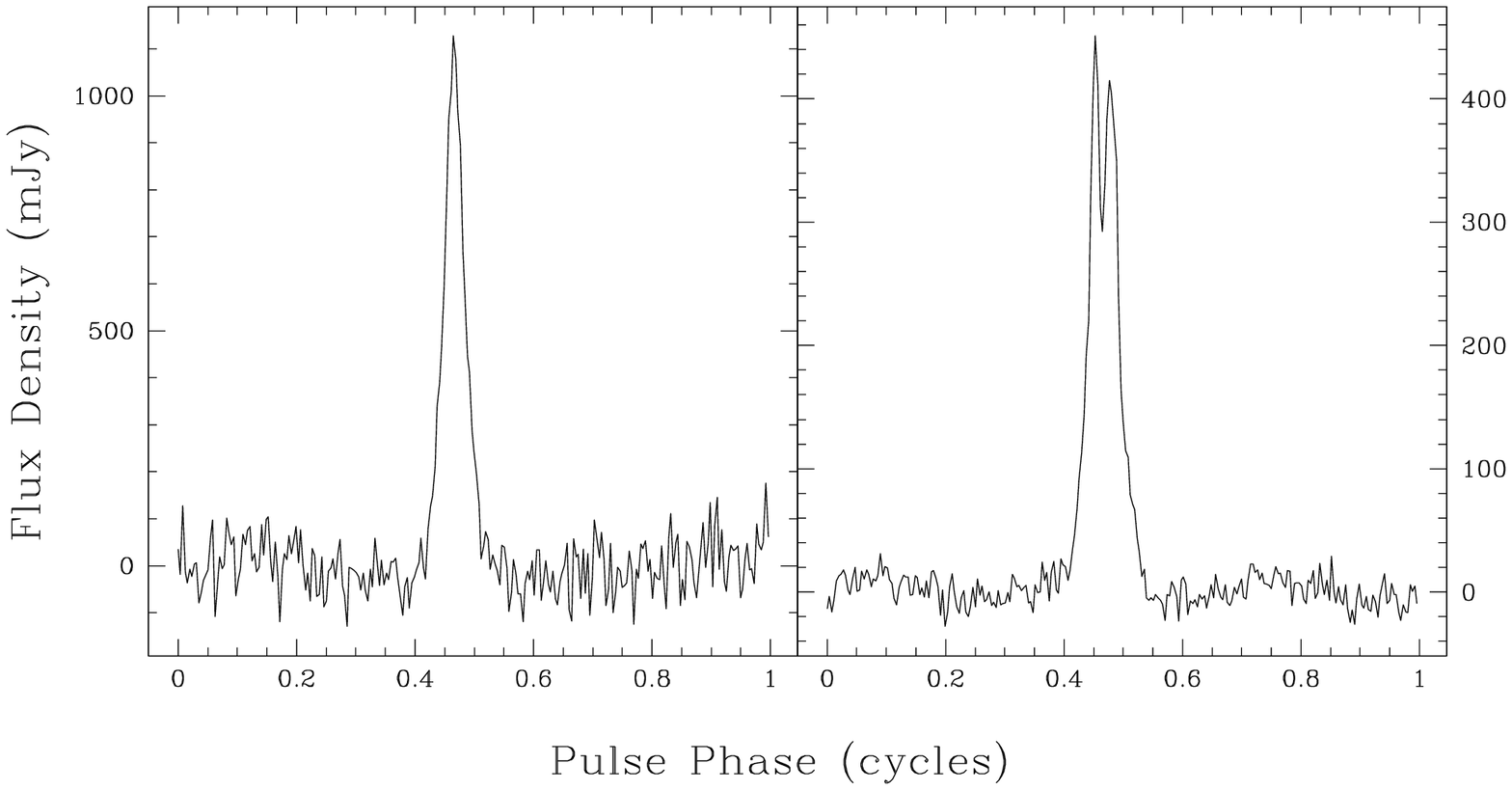}
\figcaption{
Folded pulse profiles for PSR B0355+54, which has a period of 156 ms and DM of
57 pc cm$^{-3}$, at 327 MHz
 (left) and 1435 MHz (right).
Note that these
profiles are for 5 and 2 minutes of data at 327 and 1435 MHz,
whereas our longest XPS search scans were 50 minutes
at these same frequencies.
}
\bigskip

An independent Fourier analysis of the same data, carried
out at the New Mexico Institute of Mining and Technology,
was also unsuccessful. This analysis was sensitive to
periods $\ge$10 ms and
   covered a smaller DM range, but with a finer DM spacing.
Three filter-bank channels with high levels of RFI
   were rejected, decreasing the number of spurious
   candidates significantly with only a 12\% sensitivity loss. Power spectra were
harmonically summed up to a maximum of 32 harmonics.
Final pulsar candidates were evaluated by
comparing average profiles from two halves of the bandpass and two halves of profiles
folded over a two period span. These were compared with the corresponding profiles
for non-dedispersed data.

The Crab pulsar, currently the youngest known radio pulsar,
was initially detected through its `giant' pulses, not through its time-averaged pulsed emission (\cite{sta68}).
Giant pulses are individual pulses with amplitudes much greater
than the mean pulse amplitude (\cite{lund95}). As we might expect other young pulsars to behave
similarly to the Crab, we also searched for 
aperiodic, dispersed pulses from the XPS. We recorded all
pulses
with amplitudes above a signal-to-noise threshold of 4 
 for each dedispersed time series,
enchancing our sensitivity to broadened pulses by repeating this thresholding
with different
levels
of time-series smoothing. Given our 4$\sigma$
threshold, we were sensitive to all single pulses with flux densities greater than $25/\sqrt{w_{ms}}$
Jy,
where $w_{ms}$ is the pulse width in milliseconds, at 327 MHz and $1.0/\sqrt{w_{ms}}$ Jy at 1435 MHz. Our
smoothing algorithm optimized the search for pulse widths up to 15 ms. (For comparison, the widths of the
Crab pulsar's giant pulses range from roughly 0.2 to 0.4 ms.) 
We found no evidence for isolated pulses from the XPS.
The result of this analysis, along with the result for PSR B0355+54, is shown in
Figure 2. For comparison, the Crab pulsar emits
roughly 80 pulses per minute with flux densities greater than 100 Jy at 327 MHz
and 8 pulses per minute with flux densities greater than
50 Jy at 1435 MHz.

\medskip
\epsfxsize=9truecm
\epsfbox{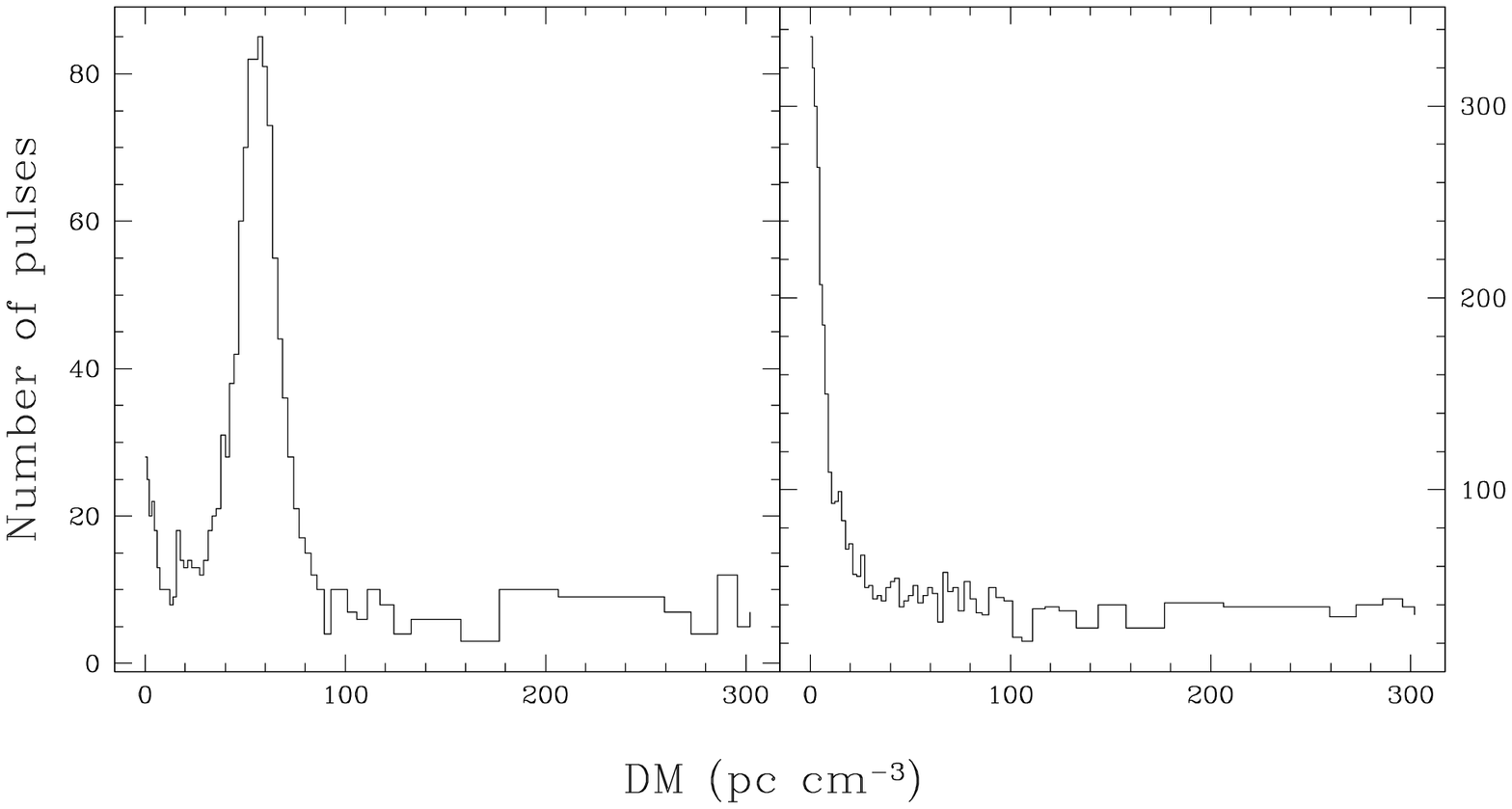}
\figcaption{
The number of isolated, dispersed pulses at 327 MHz above a $4\sigma$ threshold is
plotted vs. DM for a 5-minute observation of
PSR B0355+54 (left) and a 50-minute observation of the XPS (right).
 While a peak at the 57 pc cm$^{-3}$
DM of PSR B0355+54 is obvious, there is no evidence for such an excess
from the XPS. The peak at low DM in the right plot is consistent with interference,
and is similar to the number of counts observed in the consecutive blank sky pointing.  I
n the absence of
radio frequency interference and/or a pulsar, we would expect a flat distribution.
}
\bigskip

\section{Results}

To calculate an upper limit for pulsed radio emission from the XPS, we must account for
the excess system temperature due to the emission from Cas~A itself.
The Cas~A remnant
is $5'$ in size (\cite{green2000}). Hence, the ratios of our beam area
to remnant area are 0.003 and 0.0002 at 327 and 1435 MHz. Given Cas~A's
measured flux
density and spectral index (\cite{green2000}), we calculate remnant flux densities of
6432 and 2060 Janskys at 327 and 1435 MHz, contributing to an increase in system temperature
of a modest 40 K at 327 MHz and a mere 1 K at 1435 MHz for the phased array. However, while the phased array
resolves out emission from the nebula, each single dish will see the whole remnant,
leading to an increase in system temperature of 475 K at 327 MHz and 230 K at 1435 MHz.
By comparison, a single dish with the
same collecting area as the VLA would suffer increases of system temperature of 13000 and 6200 K at
327 and 1435 MHz.

We calculate the minimum
detectable flux density of our search using the expression (\cite{cordes97})

\begin{equation}
S_{min} = \frac{\eta T_{sys}}{G \sqrt{N_{pol} \Delta \nu \Delta t N_{\rm FFT}}} \frac{\sqrt{N_{h}}}{\sum_{l=1}^{N_{h}} R_{l}},
\end{equation}
where $\eta$ is the signal-to-noise threshold used, $T_{sys}$ is the system temperature, $G$
is the telescope gain, $N_{pol}$ is the number of polarization channels, $\Delta \nu$ is
the total bandwidth, $\Delta t$ is the sample interval, $N_{\rm FFT}$ is the number of points in the
FFT, $N_h$ is the optimal number of harmonics with which a pulsar can be detected, and
$R_{l} \le 1$ is the ratio of the $l$th harmonic to the fundamental. $N_h$
will
depend on the pulsar's duty cycle, or ratio of pulse width to period,
 with narrower pulses producing more harmonics. For this calculation,
we take the intrinsic duty cycle to be the minimum of 0.3 and $0.03(P_{ms}/1000.)^{-0.5}$, 
where $P_{ms}$ is the pulse period in milliseconds (\cite{biggs90}).
 Pulses may then be additionally broadened by
propagation effects,
including scattering and dispersion. The Taylor \& Cordes (1993) model predicts 1.7 ms and 3 $\mu$s 
of pulse broadening
due to scattering in the direction of the XPS and for a distance of 3.4 kpc at 327 and 1435 MHz, respectively.

Figure 3 shows the dependence of our search sensitivity
on DM and period at the two observing frequencies, accounting for the effects of scattering and dispersion 
on the intrinsic duty cycle of the pulsar. Assuming values
for period and DM of 30 ms and 75 pc cm$^{-3}$,
we calculate $7\sigma$ upper limits of 30 and 1.3 mJy at 327 and 1435 MHz. As shown in Figure 3,
these limits 
will decrease for longer period, smaller duty cycle, and lower DM. There are no hardware
limitations to our maximum detectable period, determined solely by the length of the FFT.

While there is no infrared or optical evidence for
a binary companion (\cite{van86}), acceleration effects from such a companion
would cause pulse smearing, decreasing our
sensitivity. We estimate the signal-to-noise degradation introduced for various orbital systems by
taking a Gaussian pulse and applying orbital smearing by iteratively calculating emission
times across the orbit. Smearing times due to time constant, DM, and scattering are calculated
and applied to the  
orbital-smeared pulse. We calculate the best harmonic sum for the resultant waveform and compare this to
the best harmonic sum for a waveform with no orbital smearing to calculate the loss in signal-to-noise.
For a 30 ms period and duty cycle of 0.2, a 1 solar mass companion in an 8 hour circular orbit
with an
inclination angle (i.e. angle between the orbital plane and our line of sight) of $30^{\circ}$,
our sensitivity decreases by roughly 
a factor of 2, averaging over all orbital phases, for a 16 Mpoint FFT.
In the case of our 8 and 4 Mpoint FFTs, these same binary parameters would decrease our
search sensitivity by factors of 2.4 and 2.8 over the original 16 Mpoint sensitivity to an unaccelerated
pulsar. Although the nominal sensitivities are less for these shorter FFT lengths,
the shorter FFTs would
allow us to detect pulsars in fast binaries where the binary smearing during a longer FFT 
is comparable to or greater
than the pulse period.
The sensitivity would be further decreased for
smaller spin periods, faster orbits, and more massive
companions. If the orbit was eccentric, the net smearing, averaged over all orbital phases,
would be less, and the resultant sensitivity would be higher.
Accretion from a binary companion could completely quench the radio emission.

\medskip
\epsfxsize=9truecm
\epsfbox{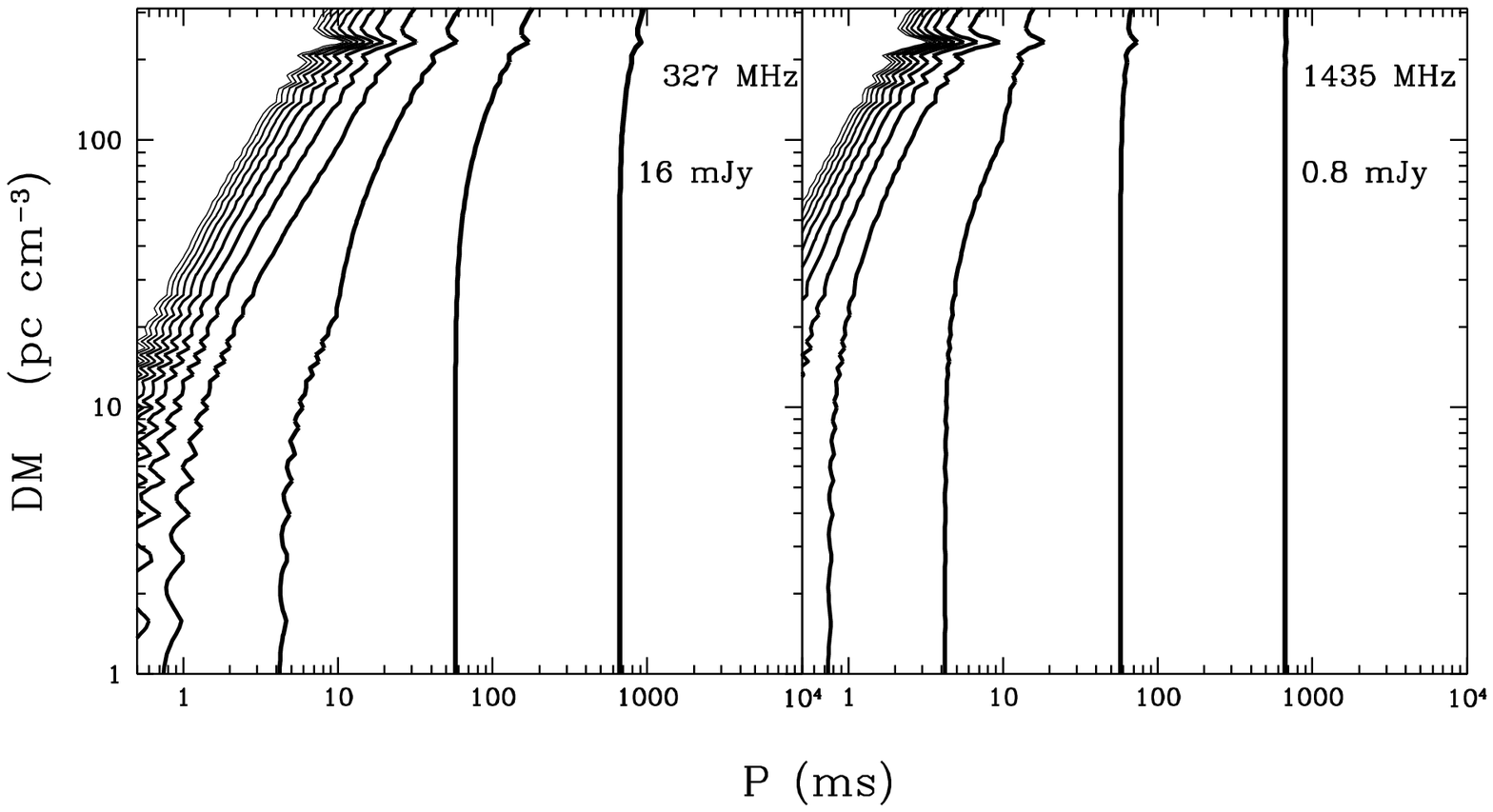}
\figcaption{
Contours of pulsed search $7\sigma$ sensitivity vs. P and DM for searches at
both 327 and 1435 MHz. The thickest (rightmost) line corresponds to a sensitivity
of 16 mJy at 327 MHz and of
0.8 mJy at 1435 MHz. Minimum detectable flux densities increase by a factor of 2 with eac
h subsequently
thinner line.
The contours are not smooth because of the finite spacing of our trial DMs.
}
\bigskip

Interstellar scintillation modulations of
the flux of the XPS are mostly negligible during our
observations. The predicted scintillation bandwidth and timescale (see Cordes \& Rickett 1998),
assuming
a uniform medium and a velocity of 50 km s$^{-1}$ are approximately 0.1 kHz and 30
seconds at 327 MHz and
0.6 MHz
and 150 seconds at 1435 MHz, in both cases too small to significantly affect our quoted upper limits.
If this velocity is higher than 50 km s$^{-1}$, as a young pulsar's velocity is likely to be (\cite{cor98}),
the predicted scintillation timescale will be smaller and will therefore be even less of a
factor in our analysis.

We have also used VLA images of Cas~A to
search for a radio continuum point-source at the position of the XPS.
At 1.3~GHz and 4.4~GHz, we find 5$\sigma$ upper limits on the flux
density from the XPS of 40~mJy and 6~mJy respectively. While not as
constraining as those derived from our VLA and Arecibo pulsed searches
described above, these continuum upper limits are completely
independent of the period, dispersion and scattering of any radio
pulsations from this source.

\section{Conclusions}
 
We have calculated upper limits of 30 and 1.3 mJy at 327 and 1435 MHz for the phase-averaged pulsed flux 
density from the XPS in Cas~A. We did not detect any single, dispersed pulses above our sensitivity
of 25 and 1.0 Jy (assuming a pulse width of 1~ms) at 327 and 1435 MHz.
We now compare our upper limit to those from previous Cas~A searches.
Davies \& Large (1970)
carried out an unsuccessful 408-MHz search for single pulses from Cas~A. 
Seiradakis \& Graham (1980) quoted a 1420-MHz upper limit of 1.6 mJy for pulsed emission from Cas~A.
However, their limit does not account for the additional system temperature contributed by the nebula,
significant for their beamwidth of $9'$.
More recently, Woan \& Duffett-Smith (1993) published an upper limit of 80 mJy at 408 MHz and       
Lorimer et al. (1998) published an upper limit of 46 mJy
at 606 MHz. For a typical pulsar spectrum, our 1435-MHz upper limit is an order
of magnitude better than these previously published limits, due to the known position of the XPS and
the power of the VLA to resolve
out the excess emission from the nebula. 

Our upper limit of 1.3 mJy at 1435 MHz translates to a luminosity of
15 mJy kpc$^{2}$ for a neutron star at 
a distance of 3.4 kpc.
This is lower than the mean 1400-MHz luminosity of 30 mJy kpc$^{2}$  estimated for young pulsars 
(\cite{frail93}) and is lower than the luminosity of any known radio pulsar
with characteristic age less than 10$^{4}$ years (see Figures 1 and 2 of Brazier \& Johnston 1999).
This suggests that the XPS in Cas~A is either not a pulsar, is a
radio-quiet pulsar, or is a radio pulsar beamed away from us.
While evidence suggests that pulsar beams are wider at short periods (see e.g.~Biggs (1990)) and that they are
therefore likely to intersect the Earth's line of sight, individual pulsars show mixed
properties. For instance, the Crab pulsar shows narrow radio components that may be part of a large
beam, but 
the profile of another young pulsar in the Large Magellanic Cloud,
PSR B0540-69, is very broad (\cite{man93}).
Furthermore, it is difficult to disentangle true beam width
from viewing angle for these pulsars. 
It is certainly possible that, like the Geminga pulsar, the XPS shows no radio
pulses but is an X-ray, and perhaps gamma-ray, pulsar. Although there is no EGRET point source
near the position of the XPS, gamma-ray searches with future, more sensitive instruments,
could be successful.
Further measurements of the spectrum and time variability of the XPS  
will also help determine which of the above scenarios is the case.

This search adds to the growing body of evidence that
there are manifestations of young neutron stars other than the standard Crab-like radio
pulsar. Although it is widely accepted that radio pulsars are born in supernovae,
fewer than 10\% of searches targeted at supernova remnants have found young (age $<$ 25 kyr) pulsars
(\cite{kaspi96,gorham96,lorimer98}).
Some of this may be due to scattering at low frequencies in the Galactic plane and the
decreased sensitivity due to the bright remnants, but it also hints at the unrecognized diversity of
the neutron star population. While some neutron stars are detected as radio pulsars, others may be born as
anomalous X-ray pulsars, soft gamma-ray repeaters, quiescent neutron stars, or other classes of objects
not yet discovered (\cite{kaspi2000,gott00}). The pathways to and percentages of these different classes of objects are not
yet clear. However, future {\it Chandra} detections of compact X-ray sources, 
and large-scale radio pulsar surveys, like the Parkes Multibeam Survey (\cite{lyne2000}), will help to answer
these questions.

\acknowledgments

We wish to thank Miller Goss for expediting our
observations and Duncan Lorimer for helpful comments on the manuscript.
We thank Larry Rudnick and Barron Koralesky
for kindly providing their radio images of Cas~A for our analysis.
This work was partially supported by NSF grant AST-9618408.
The work was also supported by the National Astronomy and Ionosphere Center,
which is
operated by Cornell University under cooperative agreement with the National
Science
Foundation (NSF). The National Radio Astronomy Observatory is a
 facility of the National Science Foundation operated under cooperative
agreement by
 Associated Universities, Inc. BMG acknowledges the support of NASA through Hubble Fellowship grant
HST-HF-01107.01-A awarded by the Space Telescope Science Institute,
which is operated by the Association of Universities for Research in
Astronomy, Inc., for NASA under contract NAS 5--26555.
VMK acknowledges support
from an Alfred P. Sloan research fellowship, from an
NSF CAREER award (AST-9875897) and an NSERC Research
Grant (RGPIN228738-00).

{}

\end{document}